# Failure of refractory masonry material under monotonic and cyclic loading – crack propagation analysis

K. Andreev[1,2], Y. Yin[2], B. Luchini[1], I. Sabirov[3]
1 – Ceramics Research Centre, Tata Steel, 1951MD, Velsen Noord, The Netherlands
2 – The State Key Laboratory of Refractories and Metallurgy, Wuhan University of Science and Technology, 430081, Wuhan, China
3 – IMDEA Materials Institute, Getafe, 28906, Madrid, Spain

**Graphical abstract**

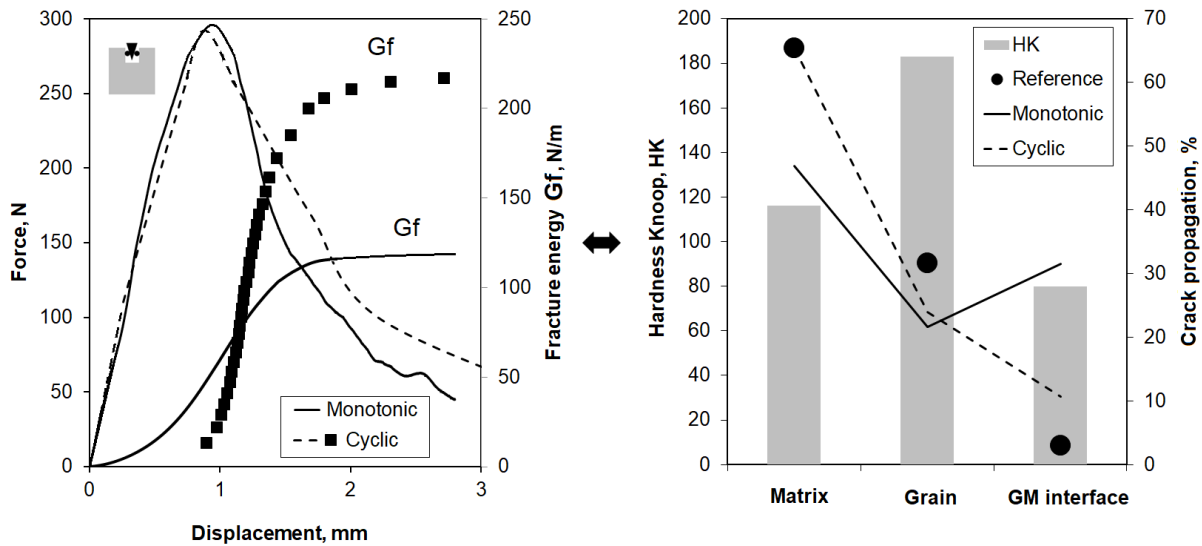

**Highlights**
- Knoop hardness interprets microstructural measurements and stress-strain data.
- Monotonic failure is more brittle, with same trans-granular failure in both modes.
- Monotonic crack aligns with less energy consuming grain-matrix interface.
- Differences arise due to lower energy input and matrix pre-cracking in the cyclic mode.

**Abstract**
Refractory masonry (refractories) is exposed to in-service loads of different types. To rationalise the masonry design and failure analysis, differences of failure under cyclic and monotonic loading were studied. For samples of silica refractories tested in wedge splitting set-up global failure parameters and crack trajectories were assessed. Under cyclic loading, higher fracture energy and lower brittleness at failure were seen. Cracks of different modes had similar non-linearity and branching. However, the size and microstructural characteristics of the fracture process zone was different. In addition, higher energy dissipation during cyclic loading is promoted by repetitive friction events along the crack trajectory.

## 1.0 Introduction
Refractory materials (refractories) such as bricks and castable concrete are used to construct industrial furnaces and high temperature reactors [1]. Refractories of different chemical and mineralogical compositions exist to accommodate the variety of processes they are utilised to contain. In service, refractories experience thermal and mechanical loads of various intensity [2,3,4]. Such loads often have a cyclic nature. For example, a typical campaign of coke making ovens features approximately 13000–15000 production cycles [5,6]. Failure of the masonry results either from a single event of an irregular spike load or from gradual degradation introduced by cyclic loads [3,7,8,9]. In the latter case, fatigue processes occur.

For the assessment of refractories, one relies on either tests simulating the service conditions or on mechanical stress–strain measurements. Thermal shock tests are an example of the former approach [5,7,8,10]. The stress–strain tests involve either monotonic [2,7,11] or cyclic loading [3,7,9,12]. Microstructural aspects of failure have been studied for individual loading modes in previous studies. Under monotonic loading, higher roughness of the fracture surface is observed for materials with higher resistance to crack propagation [13]. In materials showing more brittle failure, higher percentage of trans granular cracking is observed [11]. For cyclic stress [9,12,14] and strain [3,7,12] controlled fatigue, the deterioration of interlocking in the crack tip wake and the relocation of larger grains are believed to play a critical role in the failure. No systematic understanding of differences between cyclic and monotonic failure has been achieved so far, as reported in the literature. The comparison of cyclic strain-controlled failure and monotonic failure in specific type of refractories featuring grains of hercynite spinel demonstrated lower fracture energy for the cyclic failure [3]. Respectively, smaller fracture process zone (FPZ) was seen to develop during the cyclic loading. Contrary to that, less brittle failure during cyclic tests was seen in silica refractories tested in uni-axial compression and three-point bending [12]. For the latter, the difference was especially impressive, as monotonic loading tests showed abrupt failure and cyclic loading tests showed gradual strain softening. No microstructural analysis was performed to explain the phenomenon.

Due to the similar microstructure of matrix consisting of smaller grains, larger grains, pores and micro-cracks, refractories exhibit largely similar behaviour with regard to civil engineering concrete and rocks. For both concrete and rocks, qualitative and quantitative differences in crack propagation under monotonic and cyclic loading has been reported [15,16]. For rocks, deviations from monotonic failure trends were observed for stress and strain ("damage") controlled cyclic fatigue [15]. For concrete, whether the force-displacement curves of cyclic fatigue failure exceed or fall within the envelope of monotonic loading is assumed to be determined by the combined effect of loading amplitude and frequency, by the concrete's strength and by its heterogeneity [16]. Generally, the cyclic failure is distinguished by higher capacity of energy absorption [17] resulting from wider FPZ with crack branching [15,17]. The failure under monotonic loading is more brittle and favours trans-granular crack propagation [15,18]. The cyclic fatigue mechanisms feature de-cohesion of the larger grains and matrix loosening [15]. Final fatigue failure involves coalescence of micro-cracks [15,16]. The same phenomenon controls the saturation of the microstructural damage during repetitive thermal shock in technical ceramics [19].

In the present work, crack propagation under monotonic and cyclic loading was compared. Monotonic loading is relevant for the in-service failures due to a spike load. Cyclic loading is relevant for gradual degradation during multiple process cycles. Samples of silica brick were tested. Silica bricks are standardly used in the furnaces of glass, coke, and iron making industries [1, 5, 6,7]. Wedge splitting tests were performed. The cyclic tests followed strain-controlled protocols. Microstructural analysis was performed to explain potential differences in the mechanisms of failure under different loading modes. The geometry of the crack path was quantified by roughness and non-linearity indexes. Fractions of the crack trajectory occupied by different constituents of the microstructure were measured to indicate preferential routes of the crack propagation. Knoop micro hardness tests assessed the properties of the constituents to explain the possible preferential propagation routes. Correlations among microstructural features and the stress–strain parameters were analysed. These analyses are expected to assist the comprehension of the failure mechanisms and systematise the findings. The results of the work should contribute to the optimization of the material design and the material selection procedures.

**2.0 Methods and materials**

## 2.1 Materials

Commercially available silica bricks standardly used for construction of coke ovens were studied (Fig. 1). The bricks are pressed from quartzite grains using calcium hydroxide solution and sintered at temperatures of 1500°C. The density of the aqueous $Ca(OH)_2$ solution was 1200–1320 kg/m$^3$. The amount added was 7–8 wt.%. The chemical and physical properties of the studied silica brick were broadly discussed in a previous publication, where it was referred to as SB2 [7]. Typical chemical composition features 96.0 wt.% of $SiO_2$, 2.5 % of CaO, 1.0 % of $Al_2O_3$, and 0.5 % of $Fe_2O_3$. The mineralogical composition features 63–66 % of tridymite, 28–31 % of cristobalite, less than 1 % of quartz, and 3–5 % of pseudo-wollastonite. Tridymite forms the grains of the matrix and the rims of larger grains [7]. Cristobalite is predominantly found in the large grains (Fig. 1). In Fig. 1, the dark fields and lines are pores and their boundaries in the matrix, respectively. The grain size distribution, as indicated by the supplier, is as follows: 1–2 mm grains are 2 wt.%, 0.5–1 mm grains make up 13 wt.%, and 0.1–0.5 mm grains comprise 29 wt.%; the rest of the grains are finer than 0.088 mm. The bulk and true density of the brick are 1.84 g/cm$^3$ and 2.35 g/cm$^3$, respectively. The apparent porosity is 19.6 %. The pore size distribution obtained by Hg-porosimetry shows three peaks corresponding with the pore diameters of 0.5 µm, 10 µm, and >100 µm. The compressive strength at RT and at 1400 °C is 40–50 MPa and 10–15 MPa, respectively.

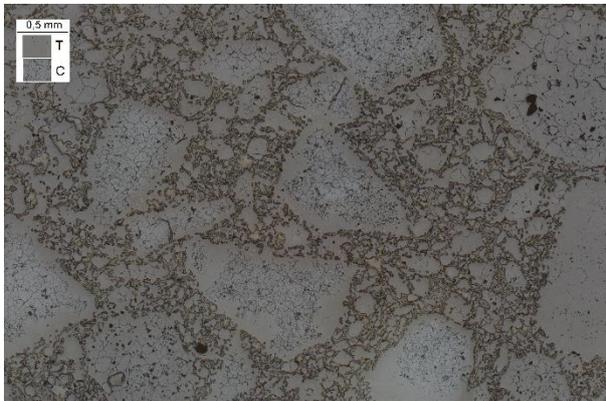

Fig. 1 Typical microstructure of the studied material: C is cristobalite, T is tridymite.

## 2.2 Methods

Monotonic and cyclic wedge splitting tests (WST) [20] were performed at room temperature (Fig. 2). WST set-up was used as it allows stable crack propagation and enables the development of the FPZ [21]. WST is closely related to the compact tension (CT) test. In both test set-ups, the stable crack propagation is promoted by the low ratio of the sample volume to the surface of the crack. The test procedure and the data analysis algorithms were discussed in previous publications [8, 22]. Material properties calculated from the force-displacement parameters of WST are notched strength (SIG-NT), fracture energy ($G_f$) and their ratio [22]. $G_f$ and strength were calculated from horizontal forces and displacements. Gf is calculated as the ratio of the surface under the force-displacement curve to the projection of the crack surface. The ratio of $G_f$/SIG-NT is the measure of brittleness at failure [11, 22]. The higher the ratio, the lower the brittleness. Refractories of low brittleness are often referred to as "flexible". WST were done in the frame by Shen Zhen Wance Testing Machine Co. The load cell capacity is 100 kN. The error of the force measurement is within 0.5 N. Estimated error of the displacement measurement is 0.5 % of the measured range. The horizontal displacements were validated by mechanical extensometers. The error of the extensometer is 0.2 µm. Samples were cut with a circular blade saw. From one brick two samples could be extracted. The width, depth, and height of the samples were 100 mm, 65 mm and 100 mm, respectively. Due to notches, the failed cross-section had the width and height of 55 mm and 66 mm, respectively. Tests were conducted with the constant vertical wedge displacement rate of 0.5 mm/min. According to established practices of testing of refractories, the loading stopped at 15 % of the maximal force [8, 20, 22]. Seven samples tested in each mode are numbered with a consecutive sample number for monotonic (SM) and cyclic (SC) modes.

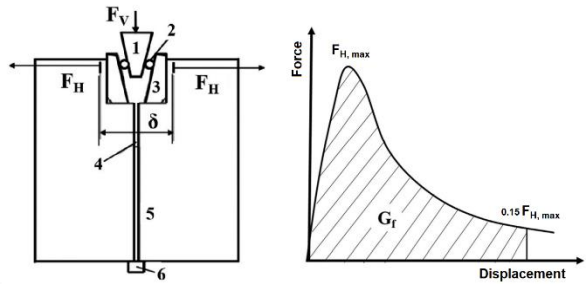

Fig. 2 Schematic representation of the wedge splitting test with a typical force-displacement curve used to calculate the fracture energy [22]: 1 – wedge, 2 – rolls, 3 – load transmission pieces, 4 - crack starting notch, 5 – side notch, 6 – support, $F_V$ – vertical force, $F_H$ – horizontal force, $\delta$- horizontal load-point displacement.

The cyclic tests were performed according to a loading protocol featuring constant effective loading displacement (Fig. 3) [12]. The loading protocol is as follows:

$$\begin{cases} D_i^T = D_{i-1}^L + \Delta D \\ D_{i-1}^L = D(F = 0) \\ \Delta D = const \end{cases} \quad (1)$$

where $D_{i-1}^L$ is the lowest displacement during the unloading in the previous (i-1) cycle, $D_i^T$ is the highest displacements during the loading in the current cycle, $\Delta D$ is the displacement amplitude. F is the force. $D_{(F=0)}$ is the displacement when force is zero. As it is shown in Formula 1 and Fig. 3, the amplitude of the loading part of the cycle is constant. The unloading stops upon reaching the minimal force level. The next loading starts where the previous unloading stopped. Such cyclic loading protocol is seen as more representative of the in-service failure of refractories, where the loading is often induced by thermal strains [12]. Due to constant displacement amplitude, the test can be seen as a fatigue test. All cyclic tests were performed with the same amplitude, which is 90 % of the average displacement at failure (displacement at the maximal force) registered in monotonic tests. Representative frequency of the test is $5*10^{-3}$ Hz. Due to the different duration of the unloading phase of the cycle, the frequency may vary through the test.

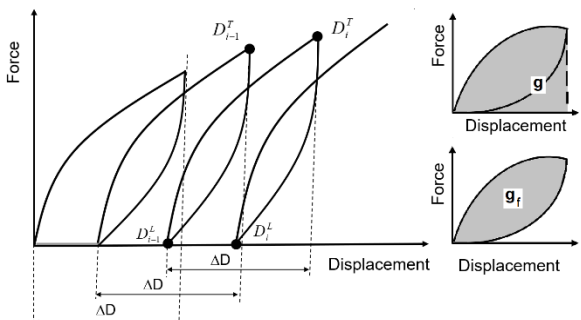

Fig. 3 Schematic representation of the cyclic loading protocol. The inserts explain the energies calculated for individual loading cycles.

For cyclic tests, the fracture energy $G_f$ was calculated from the envelope, connecting the peaks of the cycles. To monitor the build-up of the fracture energy during the loading, the parameter of partial fracture energy ($G_f^*$) is introduced. It represents the surface under the force-displacement envelope for the displacement range between 0 and some specific displacement. For the whole displacement range $G_f^* = G_f$. In addition, per cycle, the energy developed during the loading part of the cycle (g) and the energy dissipated per cycle ($g_f$) are calculated (Fig. 3, inserts). From the sum of all per cycle energies ($g_f$), the total cyclic fracture energy ($G_f^{cyc}$) was obtained. The partial cyclic energy $G_f^{cyc*}$ is the energy dissipated for the cycles from one to the current cycle. For all the cycles $G_f^{cyc*} = G_f^{cyc}$.

For microstructural analysis, it was aimed to select WST samples with maximal, minimal, and median brittleness. After the completion of WST, the whole sample was impregnated with the epoxy resin. Due to the interlocking of the cracked regions, the sample was still in one piece at the time of their impregnation by the resin. The analysis was performed on polished cross-sections cut from the sample at 25 and 50 % of the sample width, where the plane strain mode prevails. Each polished section represented the cracked part of the sample from its top to bottom. The mosaic images of the cross-sections were obtained at 5x magnification with a light optical microscope (LOM) Zeiss Axio Imager Z1 using polarized light. The software AxioVision 4.8.3. was used for the analysis of crack propagation. The crack trajectory exposed by an image was quantified by several parameters further explained in the text.

No single universal parameter of the crack geometry seems to exist. Three alternative parameters are used in this study. Differences in the physical meaning of the parameters justifies the decision. The roughness quantification was performed according to the standards ISO 11562, ISO 16610-21, ISO 4287 using the software Sigmasurf 1.0. The analysis involved the separation of the original 2D profile into waviness and roughness profiles. The spacing of points on the profile was 25 µm. The roughness parameter $R_q$ was used. It represents the root mean square of the deviation of the profile from the mean line:

$$R_q = \sqrt{\frac{1}{l_m} \int_0^m y^2(x) dx} \qquad (2)$$

where $l_m$ is the length of the profile, y(x) is the height variation. In comparison with the most widely spread roughness parameter representing the average profile height (Ra), the parameter Rq is more sensitive to the height of occasional peaks. Due to the ambiguity of the difference between the waviness and the roughness (ISO 4287), two alternative waviness cut-off values were used. Those corresponded with the wavelength values of 8 and 2.5 mm. Moreover, Fourier transform analysis was performed and the fractal Hausdorff dimension DH was calculated [22]. For the cracks, the total crack length (L2) and its projection on the crack plane (L1) were measured. L2 accounts for the crack branching. The ratio L2/L1 quantifies the non-linearity of the crack.

The crack length was measured for (i) crack propagation within the matrix, (ii) crack propagation within the large grain (trans granular crack) and (iii) the crack propagation along the grain–matrix interface. The limit between the large grain and the matrix grain is considered to be 0.1 mm. To avoid bias during measurement, the operator was not apprised of the mode of failure of the analysed sample. For reference, similar analysis was performed on random lines spanning top and bottom of the sample in the main direction of crack propagation. Two ratios characterising the crack propagation were determined. The ratio PA is the sum of lengths of cracks propagating through and along the large grains, which is divided by the total crack length L2. The ratio PBA is the ratio of length of cracks propagating through the grains in turn divided by the sum of cracks propagating through and along the large grains. Similar analysis of the crack trajectory has been performed for refractories [11, 23] and concrete [24].

Knoop hardness measurements were performed on the same WST cross-sections used for the microstructural analysis. The Knoop hardness of the large grains, the matrix, and the grain–matrix interface was determined. Per cross-section, between 10 and 15 indents were performed in each of the aforementioned three constituents to obtain averaged values specific for the sample. The established test and data analysis procedure was followed [25]. Tests with the load of 2 kg and the holding time of 15 s were done. It should be noted that in a pre-study, both Vickers and Knoop hardness tests were tried. The latter was selected for the main study as the imprint produces less auxiliary cracking upon the indentation. Moreover, Knoop imprints were easier to detect and to measure in the inhomogeneous microstructure of the silica bricks. Knoop tests with 1 kg and 2 kg produced statistically similar results for each category of microstructural constituents. Thus, the size effect was excluded in the hardness measurements. The indentation of 2 kg produced larger and easier to precisely measure imprints, especially for the matrix and grain–matrix interface.

The indentation hardness measurements are not frequently used for refractories. The information below justifies the utilization of this test in the present research. The indentation hardness is a measure of deformation resistance, which correlates with stiffness and strength [25]. For technical ceramics, the Vickers and Knoop indenters are typically used [25]. Due to its elongated shape, the Knoop intender is believed to be most suitable for brittle ceramics. The measurement accuracy is guaranteed by utilizing sufficiently high load to avoid the size effects [25,26] and to continue to prevent major cracking, which reduces the apparent hardness value [25]. Additionally, the indentation hardness tests are regularly used to study microstructure of civil engineering concrete, including interfacial transition zone between the aggregate grain and the paste of fine particles [27]. Some of such tests in macro, micro [28], and nano [29] regimes are performed on polished samples impregnated with epoxy resin. For magnesia-based refractories, Vickers hardness was used to assess the matrix in different compositions and after different heat treatment regimens [30].

## 3.0 Results
### 3.1 Mechanical tests

Typical WST force-displacement curves are presented in Fig. 4. The stiffness is approximately similar for all samples. Within both groups of samples, the strength and the strain-softening differ significantly. The samples of both loading modes show increasing brittleness (lower Gf/SIG-NT ratio) with increasing strength (Fig. 5). If the fracture energy is normalized by the ratio L2/L1, the trend becomes less significant.

The average strength is similar for SM (5.5 MPa) and SC (5.4 MPa) (Table 1). In SC, the strain softening is somewhat more gradual than in SM (Fig. 4). The fracture energy ($G_f$) and the ratio $G_f$/SIG-NT are in average 15–20 % higher in SC samples. The difference is even higher if the fracture energy obtained from individual cycles ($G_f^{cyc}$) is considered for SC. The coefficients of variation of strength and energy of both loading modes are quite similar.

In SC, the loading–unloading loops of individual cycles have an overlap. The loops are wide and are spread along displacement axis in the beginning and especially the end of the loading protocol. In the middle phase, the cycle loops are close to each other. The shape and the size of a loop is related to the energy balance of crack propagation. For individual cycles, the energy developed during the loading part of the cycle (g) and the dissipated energy ($g_f$) are shown (Fig. 4.c). In the beginning and middle phases of the test, the energy dissipated per cycle is much lower than the energy accumulated during the loading phase of the cycle. Only at the end of the loading program those values become close. At the same loading displacement, for all cycles after the first, g is lower than the envelope energy $G_f^*$. The curves representing the partial energies consumed since the beginning of the test, $G_f^*$ and $G_f^{cyc*}$, have a sigmoidal shape (Fig. 4.c). In the beginning phase of the test, the cyclic energy ($G_f^{cyc*}$) is lower than the envelope energy $G_f^*$. However, during the middle phase, a significant build-up of $G_f^{cyc*}$ occurs. On average for all cyclic samples, fracture energy calculated from the curve envelope ($G_f$) is some 30–60 % lower than that calculated from the energies dissipated in individual cycles ($G_f^{cyc}$). These findings are important for understanding the differences in the cracks of different modes. They will be treated in detail in the Discussion section of the paper.

**Table 1**
Results of WST (average for three microstructural analysis samples is in brackets).

| Sample | Cycles to failure | SIG-NT, MPa | $G_f$, J/m$^2$ | $G_f$/SIG-NT, mm$^2$/m | $G_f^{cyc}$, J/m$^2$ |
|---|---|---|---|---|---|

| Sample | | | | | |
|---|---|---|---|---|---|
| SM1 | - | 7.0 | 139 | 19.9 | - |
| SM2 | - | 5.6 | 120 | 21.4 | - |
| SM3 | - | 3.9 | 117 | 30.2 | - |
| SM4 | - | 5.8 | 114 | 19.8 | - |
| SM5 | - | 7.2 | 134 | 18.6 | - |
| SM6 | - | 5.6 | 128 | 22.8 | - |
| SM7 | - | 3.7 | 89 | 23.6 | - |
| **Average** | | **5.5 (5.6)** | **120 (125)** | **22.3 (23.8)** | |
| **CV, %** | | **24,4** | **13.4** | **17.4** | |
| SC1 | 32 | 6.0 | 115 | 19.2 | 206 |
| SC2 | 39 | 6.5 | 141 | 21.7 | 236 |
| SC3 | 115 | 3.6 | 147 | 40.4 | - |
| SC4 | 6 | 6.4 | 137 | 21.4 | - |
| SC5 | 26 | 4.3 | 139 | 32.3 | 203 |
| SC6 | 24 | 5.6 | 168 | 29.9 | 218 |
| **Average** | **40 (62)** | **5.4 (5.3)** | **143(138)** | **27.5(27.1)** | **215** |
| **CV, %** | **94.7** | **21.7** | **9.7** | **29.8** | **7.0** |
| SC7 | 15 interrupted | 3.7 | - | - | - |

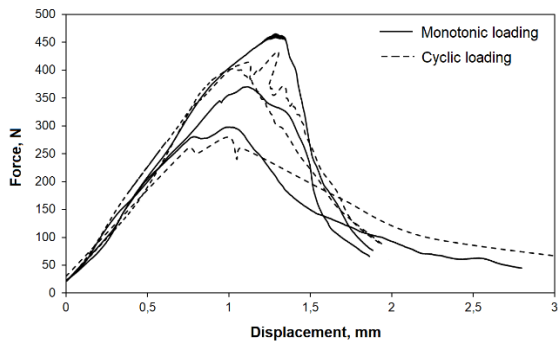

a

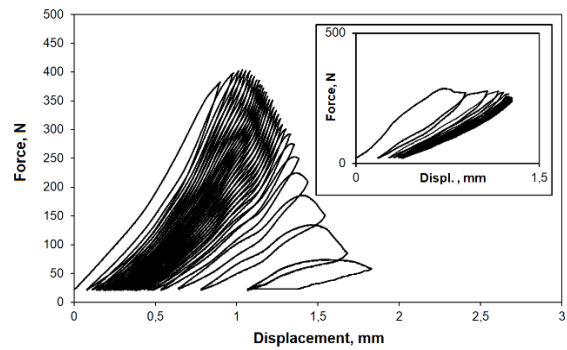

b

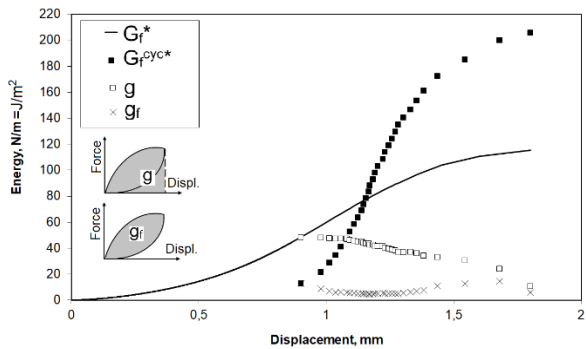

c

Fig. 4 WST curves of vertical force and displacement, (a) curves of samples subjected to microstructural analysis, (b) typical cyclic fatigue curve for SC1 and SC7 (insert), (c) energy development for the sample SC1.

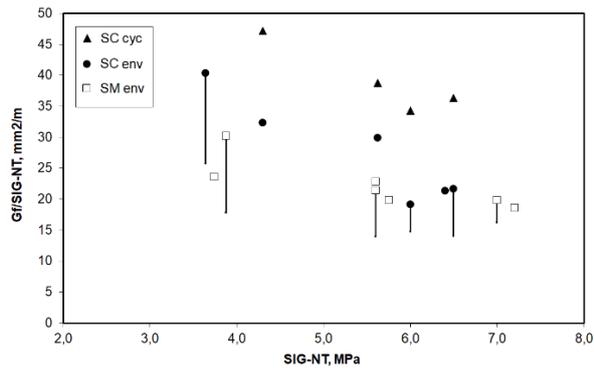

Fig. 5 Relationship between the notched strength and brittleness for individual samples. The vertical lines show the effect of the correction for L2/L1.

3.2 Microstructural analysis
The fracture in the samples 1, 2, and 3 in each group was studied by microstructural analysis (Fig. 6, 7). The crack profiles show a single major crack. From the micrographs, typical crack width is estimated to be 0.2 mm at the mouth and several micrometres at the tip. Occasional occurrences of minor crack branching are seen. In cases when an alternative crack is branched off at a higher angle, it stops quickly (e.g., SM2, SM3, Fig. 7). If the angle between the branching cracks is small, the cracks propagate for some distance almost in parallel and then re-unite (e.g., SM3, SC2). For both loading modes, the samples with cracks of highest waviness and branching (SM3, SC3, Fig. 7) are those of lowest strength (Table 1). The sample SC3 also has the lowest brittleness and needs the highest number of cycles to fail. In the crack path (Fig. 6.c, 7), there are interruptions when the crack path blends with irregularities of the microstructure. In Fig. 5, the lines with two arrowheads mark the parts of the cracks featuring frequent interruptions. In these parts the crack between the two interruptions is not significantly longer than the interruption itself. The zones of frequent crack interruptions in the tip of the crack are typically longer in SC, where they can span some 30–40 % of the crack trajectory.

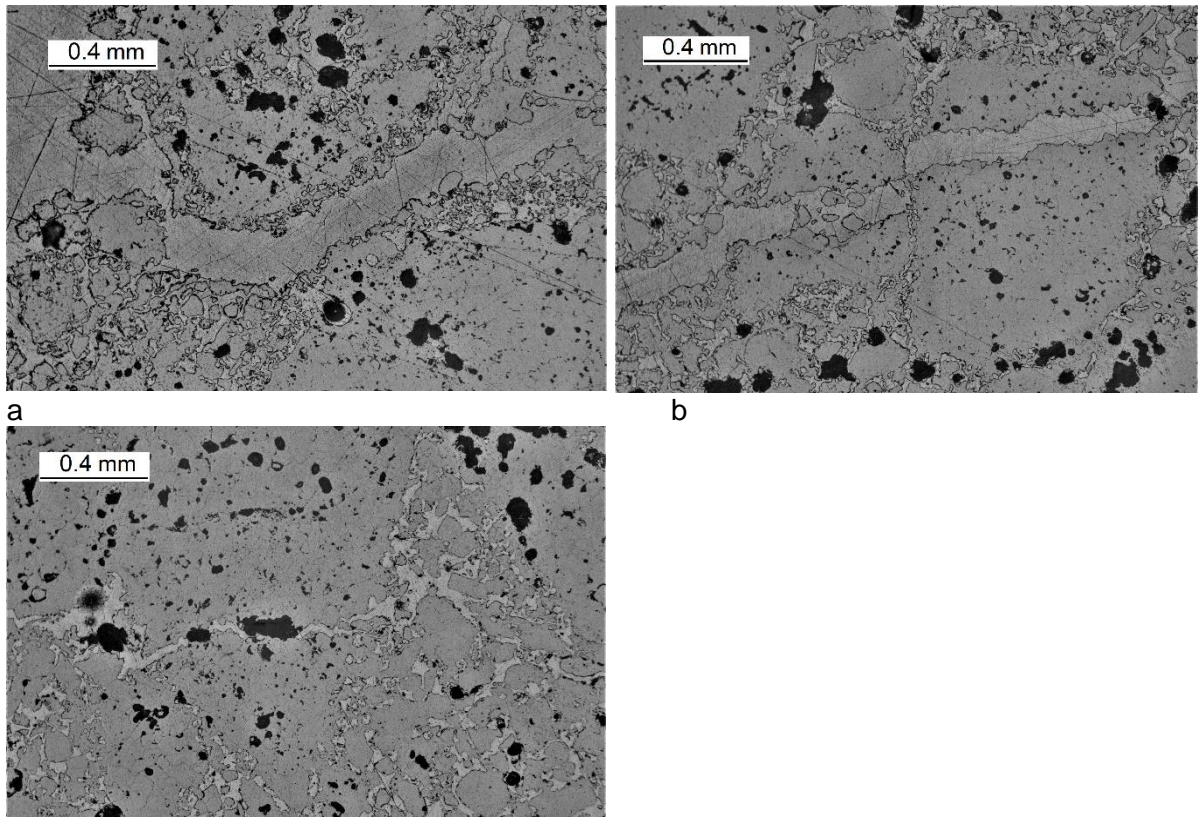

a  b

c

Fig. 6 The mouth (a), middle (b) and the tip (c) of the crack developed in SC1.

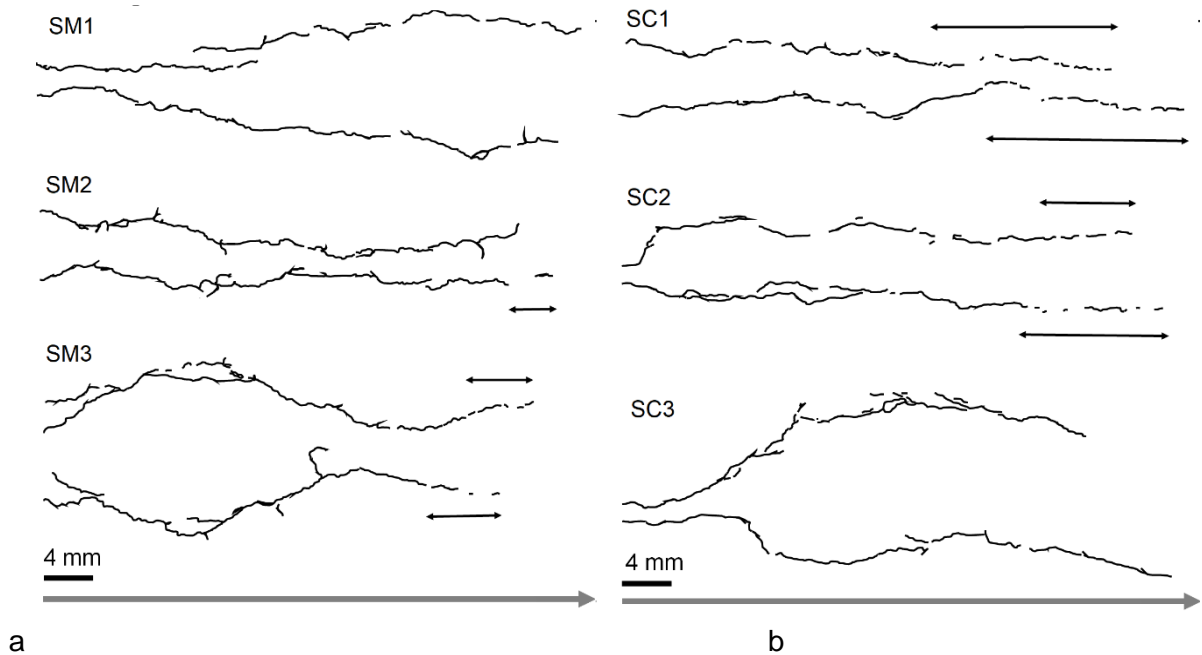

a
b

Fig. 7 Crack trajectories. The black lines with two arrowheads indicate the part of the crack featuring frequent interruptions. The grey arrow under the scale bar shows the direction of the crack propagation. Its length equals to the ligament length of the sample.

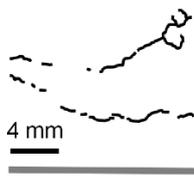

Fig. 8 Crack trajectories for SC7. The meaning of the grey arrow is as in Fig. 5.

The cross-sections of the sample SC7 (Fig. 4.b insert), whose loading was stopped after the initial 15 cycles during the middle phase of the degradation, demonstrate the presence of cracks (Fig. 8). The crack length is approximately 30–40 % of the cracks seen in the samples, which sustained full loading program. The crack path in SC7 is frequently interrupted.

The quantitative assessment of the crack trajectories indicates somewhat higher roughness for monotonic samples. The average $R_q$ for the waviness cut-off of 8 mm is 171±22 μm and 164±24 μm for SM and SC, respectively. For the waviness cut-off of 2.5 mm, it is 68±5 μm and 59±6 μm for SM and SC, respectively. The average DH number is 1.578±0.010 and 1.572±0.018. for SM and SC, respectively. The ratio L2/L1 is quite similar for SM and SC samples. The average values are 1.48±0.23 and 1.47±0.27 for monotonic and cyclic samples, respectively.

SM have higher portion of the crack propagating along the grain–matrix interface than SC and in the reference straight lines (Fig. 9). The crack portions through the grain and especially through the matrix are lower in SM. In SC, the portion of crack propagation through the matrix is as high as in the reference straight lines. The average portion of through the grain cracking in SM, SC, and in the reference straight lines is 21.6±11.2 %, 23.9±6.2 %, and 31.6±4.2 %, respectively. The average portion for through the matrix cracking in SM, SC, and in the straight lines is 46.9±7.3 %, 65.4±2.5 %, and 65.4±4.9 %, respectively.

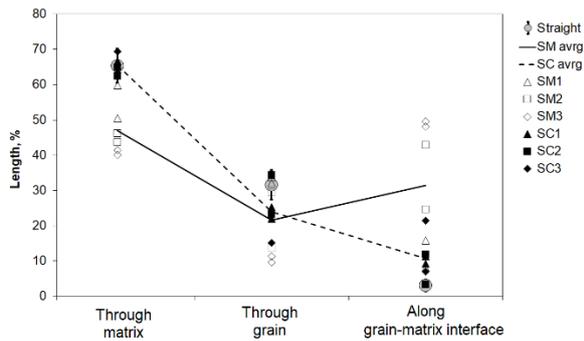

Fig. 9 Crack propagation through different constituents of the microstructure compared with the results for the random straight reference lines through un-cracked material (marked "straight").

Among the constituents of the microstructure, the hardness increases in order from interface to matrix to the large grain (Fig. 14, Table 2). The coefficient of variation for the hardness in the matrix and the interface is above 20 %. For the grain, this value is almost twice as low. The distribution of hardness in the grain is aligned with the gradient mineralogical composition featuring higher content of tridymite and cristobalite in the rims and in the centre of the grains, respectively. Hardness of all the constituents of microstructure is significantly higher than that of the epoxy resin, used to prepare the polished section. The hardness of the epoxy resin is 5 HK.

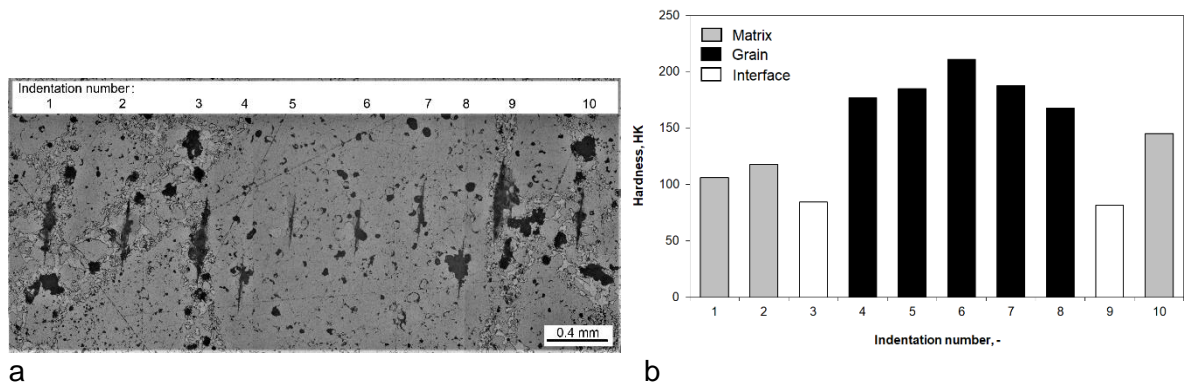

a                                   b

Fig. 10 Representative results of Knoop hardness tests in SC1, (a) indentation imprints in different microstructural constituents, (b) corresponding Knoop hardness values.

**Table 2**
Hardness data

| Sample | Matrix | Grain | Interface |
| --- | --- | --- | --- |
|  | HK (CV %) | HK (CV %) | HK (CV %) |
| SM1 | 118 (27) | 191 (7) | 86 (12) |
| SM2 | 116 (23) | 183 (12) | 80 (21) |
| SM3 | 107 (18) | 167 (13) | 65 (29) |
| SC1 | 113 (30) | 194 (15) | 85 (23) |
| SC2 | 131 (25) | 194 (13) | 84 (21) |
| SC3 | 107 (18) | 167 (13) | 65 (29) |
| SC7 | 120 (18) | 186 (13) | 92 (9) |
| **Average** | **116 (23)** | **183 (12)** | **80 (21)** |

3.3 Correlation of parameters
For individual crack trajectories, roughness parameters are compared with the crack non-linearity L2/L1 and with the microstructural indexes of crack propagation PA and PBA. Considered roughness parameters include Rq of two cut-off limits and Hausdorff dimension

DH. With L2/L1, no significant correlations are seen. The exception is the correlation between DH and L2/L1 for monotonic loading (Fig. 11). With PA and PBA, no correlations are seen for DH and $R_q$ for the cut-off of 8 mm. However, correlations are observed for Rq for the cut-off of 2.5 mm (Fig. 12). The following can be observed from those correlations. Both for PA and PBA, the data of two loading modes form common trends. For PA, the trend has an exponential form. The data of SC occupy the flatter part of the trend, where PA does not change significantly with roughness (Fig. 12.a). For PBA, a non-linear decreasing trend is observed.

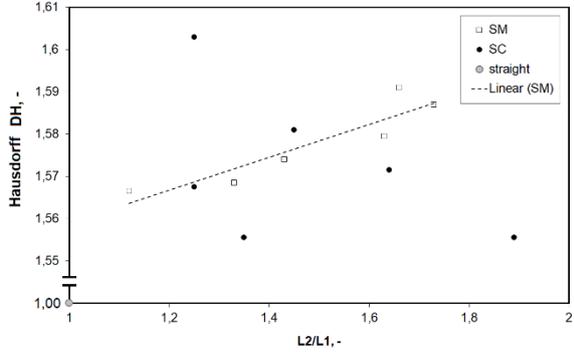

Fig. 11 Relationship between the Hausdorff constant and the ratio L2/L1.

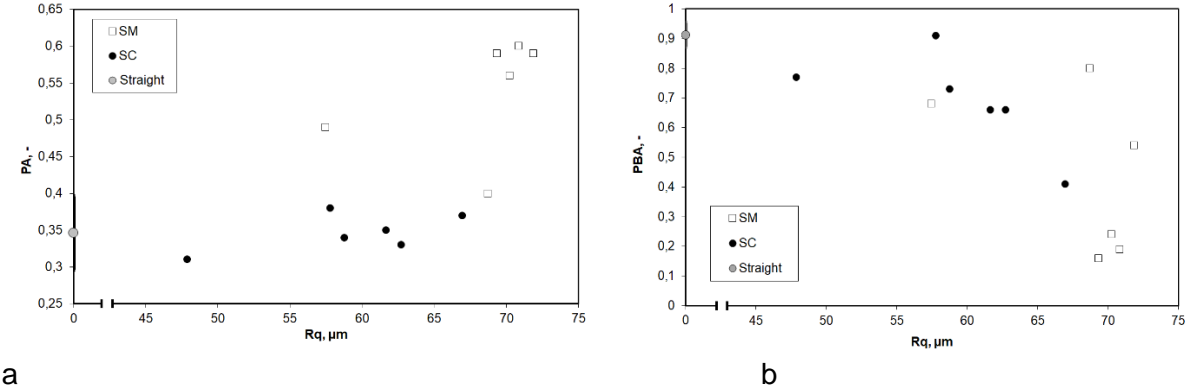

a                                                                                                          b

Fig. 12 Relationship between the ratios PA (a) and PBA (b) and the toughness Rq for the cut-off limit of 2.5 mm. "Straight" refers to the values for the reference straight lines.

When L2/L1 is compared for PA and PBA, clear trends are observed(Fig. 13). With growing L2/L1, the PA of monotonic samples linearly increases. PA of cyclic samples does not change with L2/L1. PBA of both loading modes decreases with increasing L2/L1. However, the sensitivity to the changing L2/L1 is lower for CS (flatter trend). All the trends can be extrapolated to the parameters specific for the reference straight lines (Fig. 12, 13).

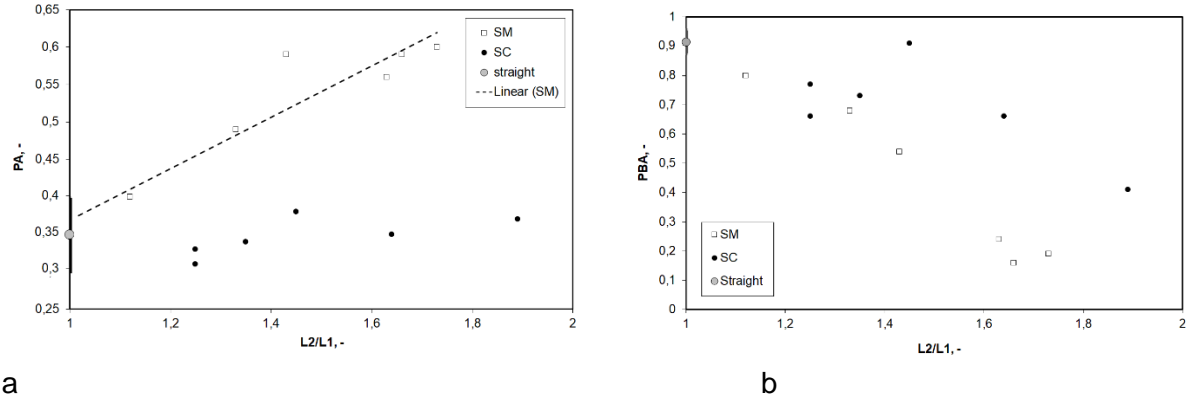

a                                                                                                          b

Fig. 13 Relationship between the ratios PA (a) and PBA (b) and the ratio L2/L1. The fit in (a) is discussed in the text.

The correlation between the stress–strain parameters and the crack trajectory specific parameters is shown in Fig. 14. Data of only three samples with known crack propagation parameters is available per loading mode. Due to this, the attempted correlations are general for both loading modes. No statistically robust correlations are seen. Among the stress–strain parameters, $G_f$ has fewer flat trends and thus is the most sensitive to the micro-structural parameters. Gf has a positive correlation with PBA. Gf has negative correlations with DH, Rq and PA. A certain relationship of falling strength with growing L2/L1 is seen. It seems that data of monotonic tests form more obvious trends than the cyclic ones.

Normalized hardness for cracks of both loading modes shows certain trends regarding all the stress–strain parameters (Fig. 14.b). The normalized hardness is obtained regarding the average hardness values of the constituents of the microstructure in the specific sample and their portion in the respective crack trajectory. For the positive correlation between the normalized hardness and SIG-NT, the coefficient of determination ($R^2$) is 0.76. This is the only correlation among those of Fig. 11 that has an $R^2$ value of above 0.7. This renders it a statistically robust correlation. As the strengths in both modes are similar, the correlation observed is attributed to the variation of the microstructural characteristics of the samples within each mode of loading. The hardness has positive and negative trends with the fracture energy and the brittleness ratio, respectively (Fig. 14.b). Normalised hardness has good correlation with PBA.

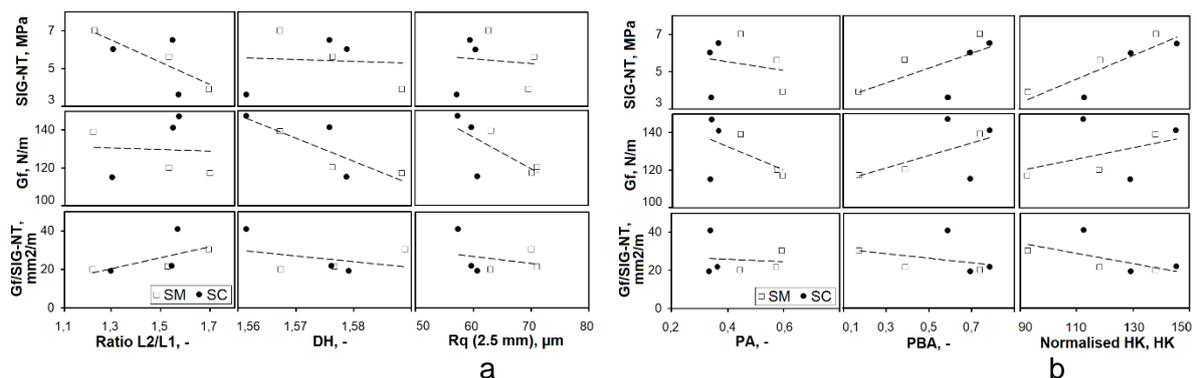

Fig. 14 Relations of the average crack trajectory parameters of a sample with its stress–strain parameters.

## 4.0 Discussion

In agreement with the results for other types of granular materials [15,17] and our previous research [12], the cyclic failure was found to be less brittle than the failure under monotonic loads. However, unlike in other materials, the reason of lower brittleness is neither more extensive crack branching (Fig. 6, 7), nor more tortuous crack path due to lower percentage of through the grain failure (Fig. 9). The crack trajectories differ predominantly by the fractions of around the grain and through the matrix failure. This aspect comprises the novelty of the reported research. The explanation of microstructural phenomena causing the differences between monotonic and cyclic failure in silica refractories is the focus of this section. First, effects of the crack propagation around the large grains are analysed. Second, the mechanisms explaining the differences in the crack trajectories and global properties are proposed. In the end, the practical importance of the findings is discussed.

### 4.1 Effects of grain–matrix interface

Roughness of the crack trajectory, especially Rq(2.5 mm), is higher in SM than in SC. This should result from higher portion of around the grain failure and be caused by the larger grains protruding on the fracture surface. The matrix consisting of smaller grains produces smother cracks. This is supported by the fact that the roughness increases with higher portion of either through or around the grain fracture (higher PA) and that it decreases with higher portion of through the grain failure (higher PBA) (Fig. 12). Smaller difference between the loading modes is seen for the roughness with the waviness cut-off threshold of 8 mm. With a higher cut-off, both waviness of lower wavelength and the roughness are counted as roughness. As the size of larger grains is 1–2 mm, the waviness cut-off threshold of 2,5 mm seems to be more appropriate.

In SM, higher portion of the around the grain failure could cause higher interlocking of the crack sides increasing the resistance to the crack propagation. However, for the studied material such effects are not prominent. Contrary to the expectations, for monotonic tests, higher roughness (Rq, DH) and higher portion of around the grain failure (PBA) tend to produce lower Gf (Fig. 14). The indentation tests show that the grain matrix boundary is weaker than both matrix and the grain. Probably, the energy needed to produce the new crack surfaces along the interface is rather low, which can outweigh the energy consumption due to the interlocking.

The weakness of the grain–matrix interface can be due to higher porosity in the interface zone and lower sintering activity of the larger grains. The former results from the densification processes while pressing the brick. Under pressure smaller grains are more mobile. Due to mutual penetration, densification on an imaginary boundary between two groups of smaller grains is much easier than on the boundary between a larger grain and the matrix. The difference in sintering activity can be related to higher number of contacts between the small grains of the matrix than between the matrix and the large grain. Moreover, larger grains have more favourable ratio of the surface to volume than the smaller grains. Thus, their sintering activity is lower.

4.2 Failure mechanisms
Within each group of samples, the strength is related to the crack non-linearity (Fig. 14.a) and to the through the grain failure (Fig. 13). Higher stress at failure typically means higher elastic energy built-up before the crack initiation. This should explain the fact that cracks in samples of higher strength have lower non-linearity L2/L1 (Fig. 14.a). Such cracks deviate less from the flat crack plane (straight fracture line) (Fig. 7) and have higher portion of the through the grain failure (Fig 13.b). The cracks propagating with less energy available are possibly less sensitive to the main cracking direction and deviate towards local defects. As an example, one should refer to SC3, which due to the deviation of the crack, sustained significantly more loading cycles than other SC samples. Higher fracture surface (higher L2/L1) of the lower strength samples also promotes energy dissipation. This reduces the brittleness even further. The illustration of the latter effect can be seen in the flatter trends between the strength and $G_f$/SIG-NT when the latter is corrected by L2/L1 (Fig. 5).

Comparing the performance of two sample groups, lower brittleness of SC cannot be explained either by different strength or by different crack non-linearity L2/L1. On average those are similar in both modes. Higher energy dissipation in SC is explained by following phenomena. SC seems to have larger FPZ. The microstructural set-up of the FPZ is different. Repetitive actions of crack opening and closing typical for SC promote the energy dissipation.

Longer fracture process zone shows itself in SC as a part of the crack featuring frequent interruptions. In Fig. 7, this part is marked by the lines with two arrowheads. The absence of branching in the process zone is due to the brittle nature of silica refractories. This results from the strong cohesion between the grains and the lack of mismatch in the properties in the constituents of the microstructure [7].

An explanation of differences in the crack trajectories and FPZ may be based on the energy balance. Fracture is determined by the balance between the elastic energy available at the onset of failure and the ability of the material to dissipate it. At failure, higher energy built-up results in faster acceleration of cracks [31]. For concrete, conflicting mechanisms can control a propagating crack [32]. According to the first mechanism, higher energy input has a potential to distribute more energy in a larger zone. Moreover, in brittle materials such as glass, higher ratio between available energy and the energy consumed by the crack formation is characterized by intensive crack branching and oscillation [31, 33]. The low energy crack is straight and has no branching. According to the second mechanism, higher energy input has a lower ability of stress relaxation at the crack tip and has lower fracture process zone. Materials with lower fracture process zone are to develop more linear cracks. In our case, the cyclic loading is characterized by lower energies available to propagate the crack (g vs $G_f^*$ in Fig. 4.c). In Fig. 4.c, the curve of $G_f^*$ is also representative for the monotonic loading. Moreover, regular unloading SC is to deaccelerate the crack. The first mechanism could be used to explain the differences between the two loading modes. Higher potential of the monotonic loading to distribute the energy and to develop oscillating cracks may explain why monotonic cracks "find" the grain matrix interfaces, which are weaker and thus more profitable for the crack propagation. The second mechanism may explain the differences between the samples of the same group. Those include less through the grain failure and higher L2/L1 in samples of lower strength. The overlapping effects of two mechanisms may occur, when due to the first mechanism, the monotonic crack meets the grain and when due to the second mechanism, it either cracks it or grows around it with subsequent deflection. In the cyclic samples, the low sensitivity of PBA to the changing L2/L1 is due to overall lower energy levels.

Apart from the energy balance, the preference for through the matrix propagation under cyclic loading can be influenced by the matrix damaging the frontal zone of the crack. Such damaging (micro-cracks) develops according to the mechanisms specific for cyclic loading and involving occurrence of local stresses due to interlocking (Fig. 6.b) and subsequent irreversible displacements. On the force-displacement curve, the latter are the displacements at the end of unloading. Rotation of larger grains under repetitive loading and subsequent cracking of the matrix was described as the mechanism initiating the cyclic fatigue failure in carbon bonded refractories [34]. Under loading, the large grain of an irregular shape can rotate. Upon unloading, friction and debris formation does not allow the grain to return to its original position. Instead, damage occurs in the matrix to re-allocate the grain. The corners of the large grains are the most probable location of the damage formation. With subsequent cycles, the degradation becomes progressive and leads to the formation of micro-cracks in the matrix. The micro-cracks formed ahead of the crack tip can effectively divert the crack from along the grain–matrix interface propagation.

During the cyclic tests, friction due to repetitive opening and closing of (micro-)cracks occurring in FPZ promotes energy dissipation. Larger FPZ further intensifies this effect. Respectively, one should notice intensive growth of $G_f^{cyc*}$ seen in the middle phase of the cyclic WST (Fig. 4.c). For cyclic fatigue tests, the middle phase of degradation is characterized by overcoming the micro-structural resistance to the crack growth, e.g., by the reduction of friction and interlocking [12,15,16,35]. Micro-damage growth and coalescence occurs in this phase. Presence of multiple non-connected cracks in the sample SC7, where the loading was stopped during the middle phase, illustrates the process. Micrographs of crack trajectories show numerous instances of interlocking. E.g. in Fig. 6.b one sees an interlocking between a large grain and matrix. During cyclic loading repetitive evens of gliding occurred at this location. Some debris present near the location are believed to indicate the effects of friction responsible for the increased energy consumption during cyclic loading.

4.3 Practical importance
The evidence of different fracture behaviour under monotonic and cyclic loading has direct importance for the assessment of materials and the design of refractories masonry. Whether the cyclic failure is characterised by lower brittleness and higher fracture energy for all classes of refractories is not clear. Regarding the complexity of the observed mechanisms and the variety of existing micro-structures in refractory masonry, the scenario when higher fracture energy failure develops during the monotonic loading seems to be possible. A complex nature of correlations between the global material failure properties and the micro-structural parameters of the crack trajectory were noted elsewhere [11]. From our study it seems that the ratio L2/L1 and the normalized hardness are most suitable for such correlations.

The analysis presented here is centred at the mechanical loads. Apart from mechanical loads, refractories are also exposed to thermal shock loads caused by the fluctuation of ambient temperature in the furnace. Regarding general similarities between the cyclic mechanical and thermal shock failure seen for refractories [36], technical ceramics [36, 37] and steel [38], effects similar to those described above can be expected for thermal shock. Thus, differences between the cracks caused either by a single thermal shock event or by multiple repetitive thermal shocks will be similar to those reported by us in this paper.

Further research should aim to demonstrate the applicability of the present finding to other classes of refractories and to the thermal shock induced failure.

## Conclusions
The paper presents the evidence of differences in stress–strain and microstructural parameters of fracture in silica refractories under monotonic and cyclic loading. The following was found:
- The cyclic loading produces less brittle failure, characterised by similar strength but higher fracture energy. Higher energy dissipation capacity is explained by differences in size and microstructural characteristics of the fracture process zone. Repetitive friction events further promote the energy dissipation.
- The crack trajectories of the two modes differ by the fractions of around the grain (along the grain-matrix interface) and through the matrix failure. The cyclic crack follows the less energy efficient through the matrix route. The inability to "find" the grain-matrix interface is explained by two factors. Firstly, under cyclic loading the energy input is lower than in the monotonic loading. Secondly, the cyclic crack is to be diverted from the around the grain propagation by micro-cracks. Those are most probably formed in the matrix due to degradation mechanisms typical of cyclic fatigue loading, e.g., large grain rotation.
- In general, the observed mechanisms are different from the known mechanisms of higher crack non-linearity and lower portion of the through the grain failure expected in the cyclic loading. However, between the samples of the same loading mode, the strength and brittleness do correlate with both the crack non-linearity and the through the grain failure.
- For the analysis of the in-service performance of refractories, it is advised to consider different fracture energy values for failure in monotonic and cyclic modes. Discovered microstructural aspects should be considered in post-mortem analysis of failure.

## Conflict of interest
The authors declare no conflict of interest.

## Acknowledgements
Part of the work done at WUST was done with the grant National Key R&D Program of China (No. 2018YFF0214500). The authors are thankful to Tata Steel Europe for in-kind support of the work. The support of Paula Luna Dias in processing of data is gratefully acknowledged.

## Appendix I.
## List of abbreviations and nomenclature

| | |
|---|---|
| CT | compact tension |
| CV | coefficient of variation |
| DH | Hausdorff dimension |
| FPZ | fracture process zone |
| SB | silica brick |
| SC | sample tested in cyclic mode |
| SM | sample tested in monotonous mode |
| WST | wedge splitting test |
| $D_{i-1}^{L}$ | lowest displacement during the unloading in the previous (i-1) cycle |
| $D_{i}^{T}$ | highest displacement during the loading in the current cycle |
| $D_{(F=0)}$ | displacement when force is zero |
| $\Delta D$ | displacement amplitude |
| F | force |
| $F_H$ | horizontal force |
| $F_V$ | vertical force |
| g | energy developed during the loading part of the cycle |
| $g_f$ | energy dissipated per cycle |
| $G_f$ | fracture energy |
| $G_f^*$ | partial fracture energy |
| $G_f^{cyc}$ | total cyclic fracture energy |
| $G_f^{cyc*}$ | total energy dissipated during cycling |
| HK | Knoop hardness |
| $l_m$ | length of the profile |
| L1 | projection of the total crack length on the crack plane |
| L2 | total crack length |
| PA | sum of lengths of cracks propagating through and along the large grains divided by the total crack length |
| PBA | ratio of length of cracks propagating through the grains divided by the sum of cracks propagating through and along the large grains |
| $R_a$ | average profile height |
| $R_q$ | roughness parameter |
| $R^2$ | coefficient of determination |
| SIG-NT | notched strength |
| y(x) | height variation |
| $\delta$ | horizontal load-point displacement |